\documentclass[a4paper]{jpconf}
\usepackage{graphicx}
\begin{document}

\title{Ion detector for Accelerator Mass Spectrometry based on low-pressure TPC with THGEM readout}

\author{A Bondar$^{1,2}$, A Buzulutskov$^{1,2}$, E Frolov$^{1,2}$, V Parkhomchuk$^{1,2}$, A Petrozhitskiy$^1$, T Shakirova$^{1,2}$, A Sokolov$^{1,2}$}

\address{$^1$ Budker Institute of Nuclear Physics SB RAS, Lavrentiev avenue 11, Novosibirsk 630090, Russia}
\address{$^2$ Novosibirsk State University, Pirogov street 2, Novosibirsk 630090, Russia}

\ead{T.M.Shakirova@inp.nsk.su}

\begin{abstract}
A new technique for ion identification in Accelerator Mass Spectrometry (AMS) has been proposed by measuring the ion track ranges using a low-pressure TPC. As a proof of principle, a low-pressure TPC with charge readout using a THGEM multiplier was developed. The tracks of alpha particles from various radioactive sources were successfully recorded in the TPC. The track ranges were measured with a high accuracy, reaching the 2\% resolution level. Using these results and the SRIM code simulation, it is shown that the isobaric boron and beryllium ions can be effectively separated at ten sigma level. It is expected that this technique will be applied in the AMS facility in Novosibirsk for dating geological objects, in particular for the geochronology of Cenozoic Era.
\end{abstract}

\section{Introduction}
Accelerator mass spectrometry (AMS) is an ultra-sensitive method of counting individual atoms, usually rare radioactive atoms with a long half-life \cite{LitherUMSA}. The archetypal example is $^{14}$C, which has a half-life of 5730 years and an abundance in living organisms of $10^{-12}$ relative to the stable $^{12}$C isotope. Using AMS, the radiocarbon age of a sample less than 50000 years old can be determined with a precision of 0.5\% in a few minutes using a sample containing a few mg of carbon or even less \cite{AMSapplications}.

\begin{figure}[h]
	\includegraphics[width=20pc]{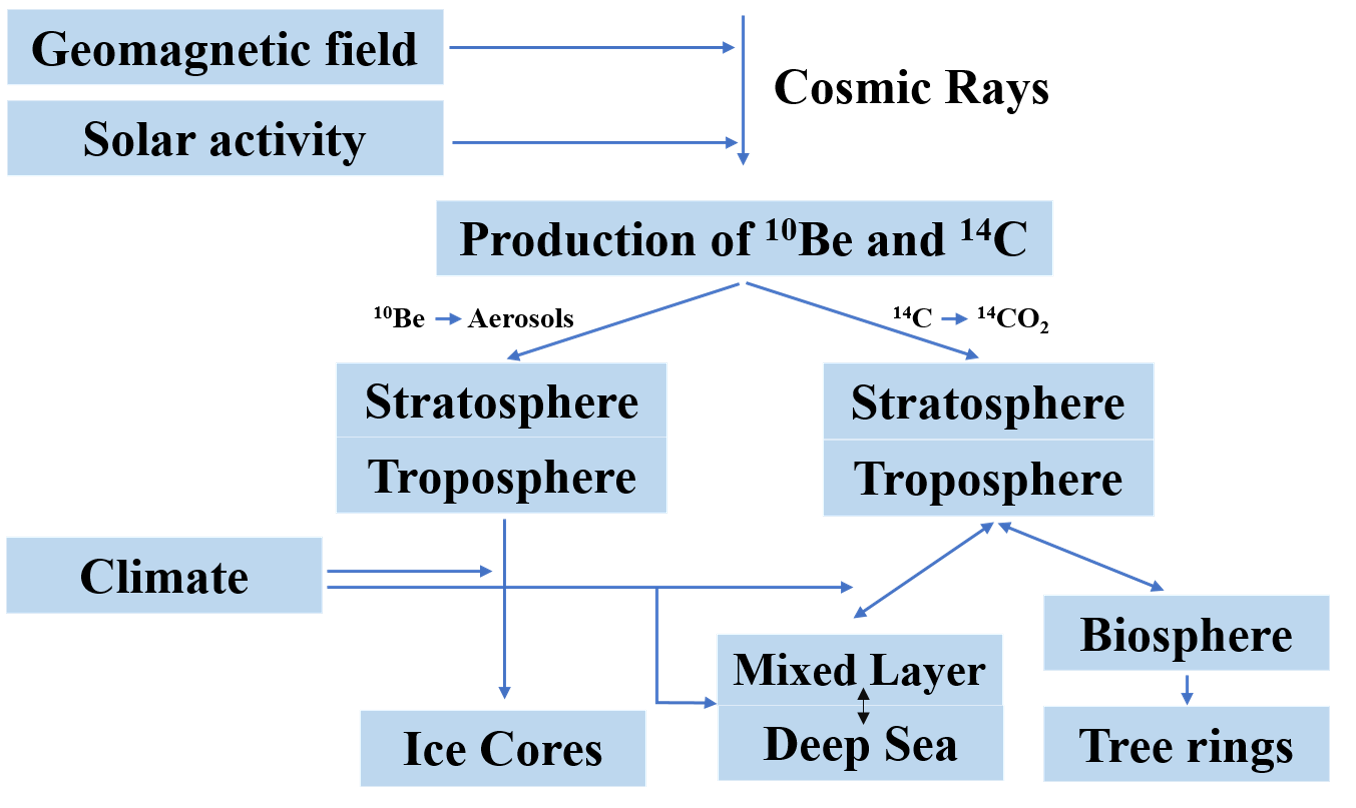}\hspace{2pc}
	\begin{minipage}[b]{14pc}
		\caption{The formation of radioactive isotopes of $^{10}$Be and $^{14}$C under the cosmic irradiation.}
		\label{Be_and_C}
	\end{minipage}
\end{figure}

The use of AMS is not limited to $^{14}$C alone, many other isotopes are amenable to the technique, the most being $^{10}$Be,  $^{26}$Al,  $^{36}$Cl and  $^{129}$I. Radioactive isotopes, their half-lives, stable isotopes and isobars are listed in the Table~\ref{tab:isotopes}. Beryllium has a longer half-life than the other isotopes, so the time interval of dating ranges between 1 thousand to 10 million years. The radioactive isotope of beryllium is produced under the constant cosmic irradiation in the atmosphere  as well as \textit{in situ} in rocks. Atmospheric beryllium is adsorbed by aerosols and then it precipitates to the earth’s surface, where it is included in various sediment. The formation of radioactive isotopes of carbon and beryllium is illustrated in the Fig.~\ref{Be_and_C}. \textit{In situ} and meteoric  $^{10}$Be are used for exposure dating to identified the growths and decays of the Antarctic ice sheet, understanding ice shelf collapse history, paleomagnetic excursions history reconstructions using ice cores, understanding the erosion rates using depth profiles of mid latitudes outcrops etc. \cite{10BeDating}.

\begin{table}[h]
	\caption{Radioactive isotopes used in AMS.}
	\label{tab:isotopes}
	\begin{center}
		\begin{tabular}{llll}
			\br
			Analyzed isotopes & Half life & Stable isotopes & Stable isobars\\
			\mr
			$^{10}$Be & 1.39 million years & $^{9}$Be & $^{10}$B\\
			$^{14}$C & 5730 years & $^{12,13}$C & $^{14}$N\\
			$^{26}$Al & 717 thousand years & $^{27}$Al & $^{26}$Mg\\
			$^{36}$Cl & 301 thousand years & $^{35,37}$Cl & $^{36}$Ar, $^{36}$S\\
			$^{41}$Ca & 102 thousand years & $^{40,42,43,44}$Ca & $^{41}$K\\
			$^{129}$I & 15.7 million years & $^{127}$I & $^{129}$Xe\\
			\br
		\end{tabular}
	\end{center}
\end{table}

AMS facilities operate in more than 100 physical laboratories worldwide, one of which is located in Novosibirsk at Geochronology of the Cenozoic Era Center for Collective Use. The principle of operation of AMS BINP is described elsewhere \cite{BINPAMS,AMS}. In the current AMS setup the time-of-flight technique is used for isotope separation. However, there is a serious problem to separate isobars, i.e. different chemical elements having the same atomic mass. The typical example are the radioactive isotopes  $^{10}$Be and $^{10}$B. To solve this problem, we propose a new technique for ion identification, namely by measuring the ion track ranges using a low-pressure time projection chamber (TPC) with Thick Gas Electron Multiplier (THGEM, \cite{THGEM}) readout.

\section{SRIM code simulation}
\begin{figure}[h]
	\includegraphics[width=20pc]{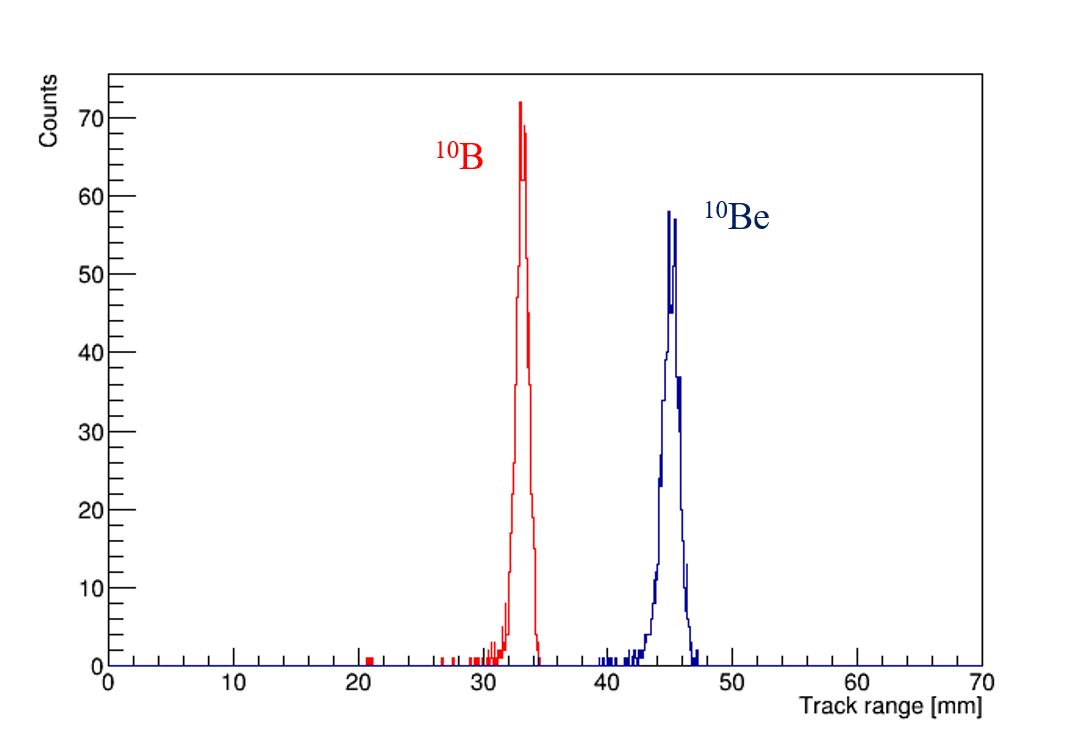}\hspace{2pc}
	\begin{minipage}[b]{14pc} 
		\caption{Track ranges distributions in the low-pressure TPC for 4.025 MeV $^{10}$B and $^{10}$Be ions for 100 nm silicon nitride window and 50 Torr isobutane, obtained using SRIM simulation.}
		\label{SRIM_B_Be}
	\end{minipage}
\end{figure}

\begin{figure}[h]
	\begin{minipage}{14pc}
		\includegraphics[width=14pc]{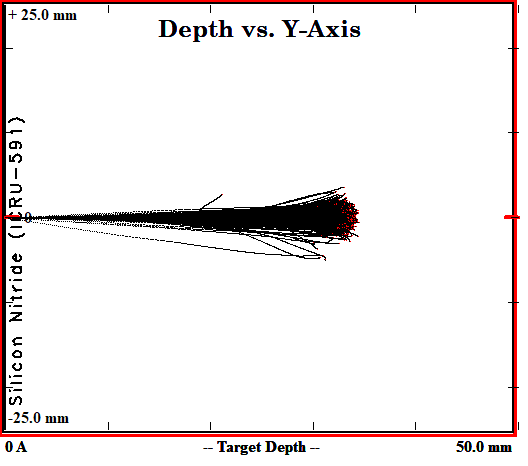}
		\caption{Tracks of $^{10}$B ions in the low-pressure TPC for 100 nm silicon nitride window and 50 Torr isobutane, obtained using SRIM simulation.}
		\label{SRIM_B}
	\end{minipage} \hspace{4pc}%
	\begin{minipage}{14pc}
	\includegraphics[width=14pc]{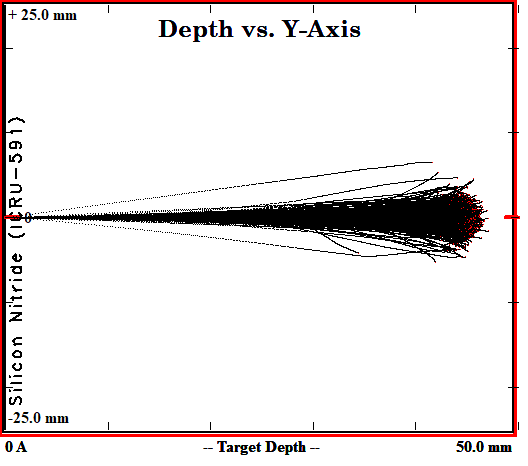}
	\caption{Tracks of $^{10}$Be ions in the low-pressure TPC for 100 nm silicon nitride window and 50 Torr isobutane, obtained using SRIM simulation.}
	\label{SRIM_Be}
	\end{minipage}	
\end{figure}

To estimate the possibility to effectively separate isobars, the SRIM simulation package \cite{SRIM} was used. In particular, the tracks of $^{10}$B and $^{10}$Be ions in low-pressure isobutane were simulated: see Fig.~\ref{SRIM_B_Be} -~\ref{SRIM_Be}. The entrance window of silicon nitride with a thickness of 100 nm was taken into account in the simulation. It can be seen from  the figures that difference in track ranges of the isobars is about 12\,mm at gas pressure of 50 Torr. Consequently, a low-pressure TPC with a spatial resolution of the order of 2 -- 3\,mm would be sufficient to separate such ions.

\section{Experimental setup}
\begin{figure}[h]
	\includegraphics[width=22pc]{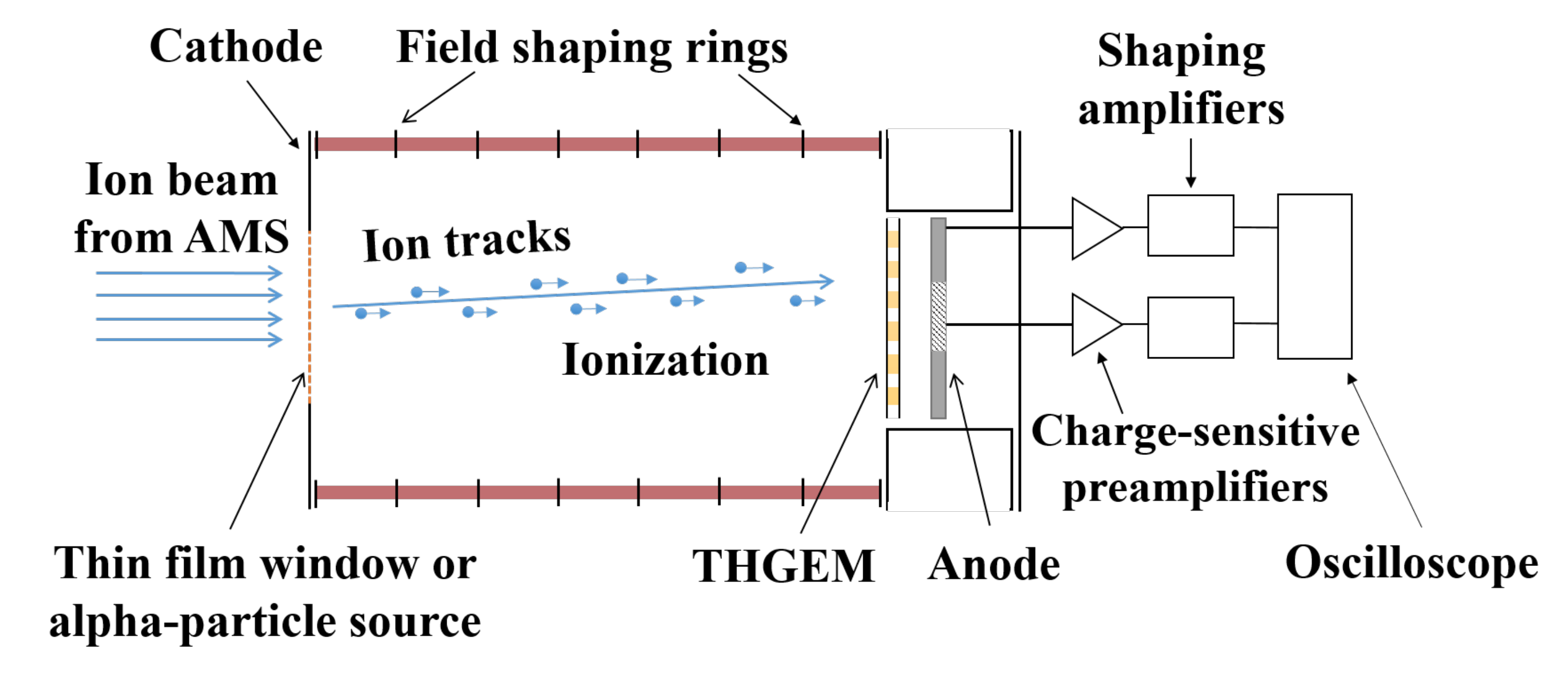}\hspace{2pc}
	\begin{minipage}[b]{12pc}
		\caption{Schematic layout of the low-pressure TPC.}
		\label{setup}
	\end{minipage}		
\end{figure}

As a proof of principle, a low-pressure TPC with charge readout using a THGEM multiplier has been developed. A schematic drawing of the detector is shown in Fig.~\ref{setup}. 

\begin{figure}[h]
	\includegraphics[width=14pc]{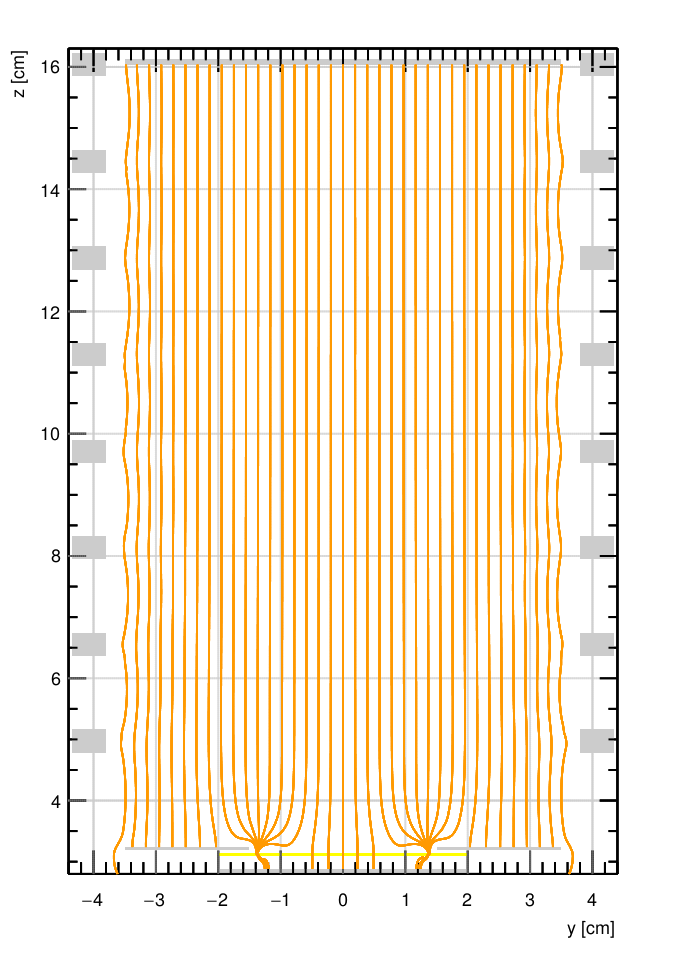}\hspace{2pc}
	\begin{minipage}[b]{18pc}
		\caption{Calculated electric field lines in the low-pressure TPC.}
		\label{el_field}
	\end{minipage}
\end{figure}

The detector comprised a cylindrical vessel made of ceramic insulator rings interleaved by the copper field shaping rings. The length of the chamber was 130 mm and the diameter was 76 mm. Thanks to the voltage applied to the field shaping rings, a uniform electric field of about 50 V/cm was formed. The Gmsh \cite{gmsh}, Elmer \cite{elmer} and Garfield++ \cite{garfield} programs were used to respectively construct the model mesh, calculate the electric field inside the chamber with FEM (Finite Element Method) and simulate electron drifting. The simulation result is shown in Fig.~\ref{el_field}. As can be seen the lines of electric fields are uniform. 

The ion beam from AMS will enter the detector volume through the membrane of silicon nitride which is used as entrance window. The membrane is produced by deposition of silicon nitride onto silicon wafers and subsequent etching from the backside. The use of silicon nitride has a number of advantages respect to other window materials, such as mylar. They are high strength, high fracture toughness, the much lower total energy loss and energy loss straggling \cite{Si3N4membranes}. For the low-pressure TPC, we will use a window with the following parameters: a frame of 14 mm x 14 mm and 525 $\mu$m thick, a membrane of 10 mm x 10 mm and 100 nm thick.
 
The ions produce the primary ionization in the detector volume. The gas pressure in the detector is adjusted in order that the ions stop near the THGEM surface. The primary ionization electrons drift in the electric field towards the THGEM with constant velocity. The amplified signal is readout at the anode. In order to select perpendicular tracks, the anode is segmented into the central disk with the diameter of 20\,mm and the guard ring with the inner and outer diameters of 24 and 30\,mm, respectively. The signals from the central circle and the guard ring are further amplified by CAEN charge sensitive pre-amplifiers A1422 followed by two NAICAM  NCB226 shaping amplifiers (the shaping time was 200 or 500 ns, the gain was 40). The amplified signals were recorded by the LeCroy 4032AR oscilloscope. Only the "central" tracks not producing a signal in the guard ring were selected for further off-line analysis.

\begin{figure}[h]
	\includegraphics[width=20pc]{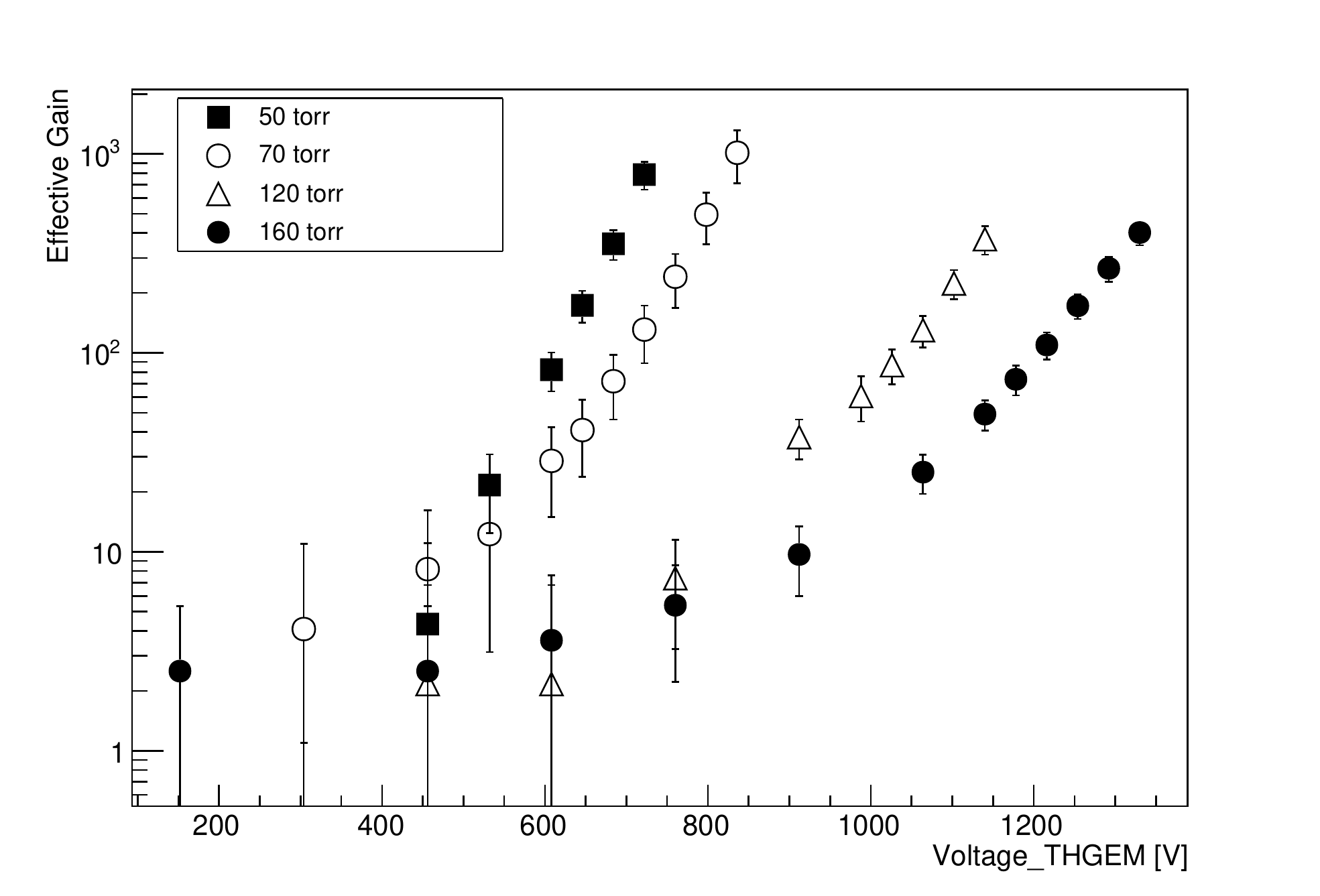}\hspace{2pc}
	\begin{minipage}[b]{14pc}
		\caption{THGEM effective gain versus the THGEM voltage bias in isobutane at pressures varying from 50 to 160 Torr in the low-pressure TPC.}
		\label{THGEM_gain}
	\end{minipage}		
\end{figure}

We studied the THGEM performance at different pressures and voltages. Fig.~\ref{THGEM_gain} depicts the effective gain curves for the THGEM at different pressures (ranging from 50 to 160 Torr). The choice of pure isobutane guarantees a reasonable compromise between a good localization of the avalanche, stable gain at low pressures and high drift velocity.

The high voltages, applied to the field shaping electrodes, THGEM and anode, were supplied by a CAEN N1470H HV programmable power supply through two independent resistive dividers. The potential on the electrodes were chosen in such a way that the zero potential was on the upper (respect to the anode) THGEM electrode. Such design allows for an independent choice of the fields in the drift and electron multiplication region of the detector.

In the laboratory, we imitate the ion beam by alpha particles of different energies: a triple alpha particle source was used combining $^{233}$U, $^{238}$Pu and $^{239}$Pu isotopes with energies of 4.8 MeV, 5.5 MeV and 5.2 MeV respectively. The alpha particle source consisted of a substrate made of stainless steel on which the active material was deposited as a thin layer. The diameter of the source active area was 11.5\,mm. The source was mounted on the upper flange, instead of the thin film window, to have a reliable electrical contact with it. To stop alpha particles of such energies within the detector volume, the isobutane pressure must be at least 120 Torr. 

\begin{figure}[h]
	\includegraphics[width=20pc]{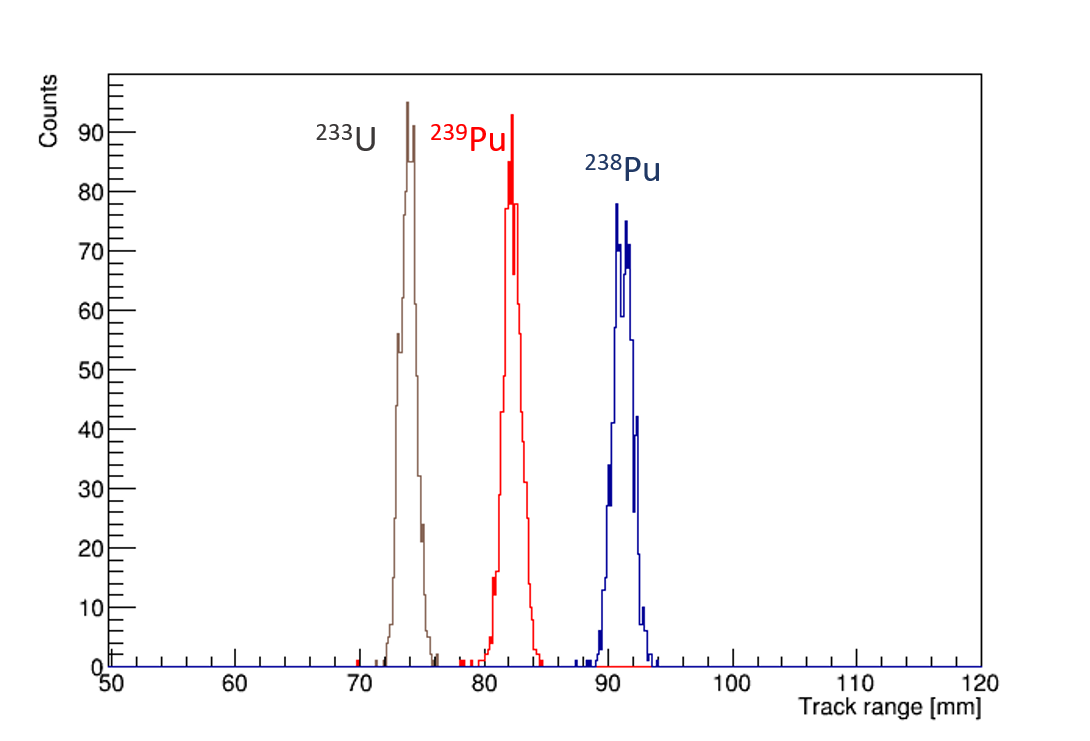}\hspace{2pc}
	\begin{minipage}[b]{14pc}
		\caption{Track ranges distributions in the low-pressure TPC for alpha particles with different energies at 120 Torr, obtained using SRIM simulation.}
		\label{TripleSourceRange}
	\end{minipage}		
\end{figure}

To make sure that the alpha particles actually stop in the gas volume, a simulation was conducted using the SRIM program (see Fig.~\ref{TripleSourceRange}).  As can be seen from the figure, the distance between the end points of the alpha particles tracks from the triple source is about 8\,mm. If we can distinguish such tracks using the low-pressure TPC, we will certainly be able to distinguish the tracks from $^{10}$B and $^{10}$Be ions in the AMS.

\section{Experimental results}
\begin{figure}[h]
	\includegraphics[width=20pc]{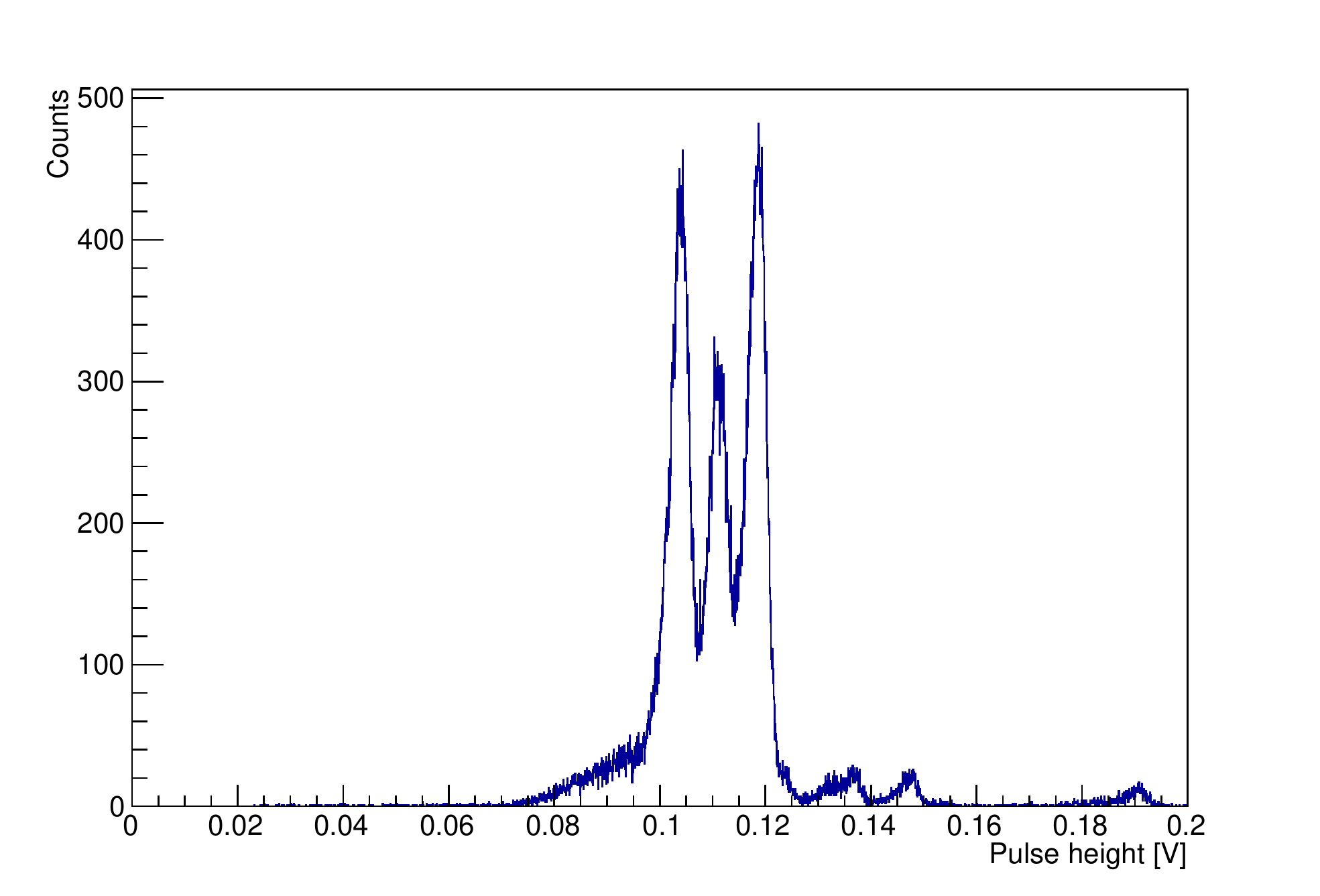}\hspace{2pc}
	\begin{minipage}[b]{14pc}
		\caption{Energy spectra of alpha particles from $^{233}$U (4.8 MeV), $^{239}$Pu (5.2 MeV) and $^{238}$Pu (5.5 MeV) sources, measured using semiconductor detector.}
		\label{TripleSpectraSD}
	\end{minipage}		
\end{figure}

The triple alpha particle source was additionally characterized using a semiconductor detector: its energy spectrum was measured in vacuum. For this exercise, the semiconductor detector was installed replacing the THGEM and the chamber was evacuated. The result is shown in Fig.~\ref{TripleSpectraSD}: the three alpha particle lines in the energy spectrum are distinctly seen with good resolution. It should be remarked that in AMS the ions cannot be identified using the semiconductor detector, since all of them have the same kinetic energy.

\begin{figure}[h]
	\includegraphics[width=20pc]{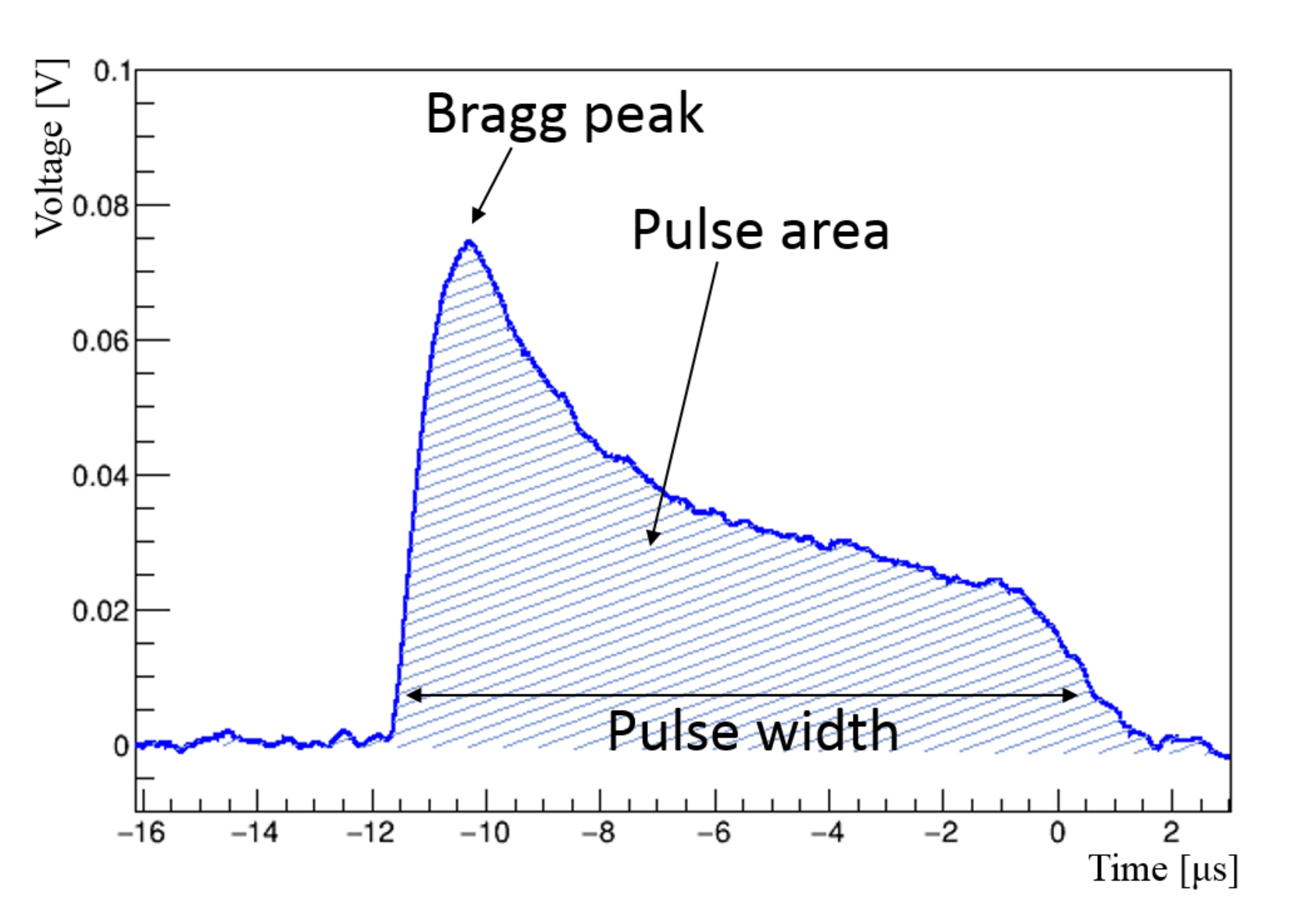}\hspace{2pc}
	\begin{minipage}[b]{14pc}
		\caption{Typical pulse shape (oscilloscope waveform) of the signal from the alpha particle in the low-pressure TPC with THGEM readout. The Bragg peak is clearly seen.}
		\label{Signal}
	\end{minipage}		
\end{figure}

\begin{figure}[h]
	\includegraphics[width=20pc]{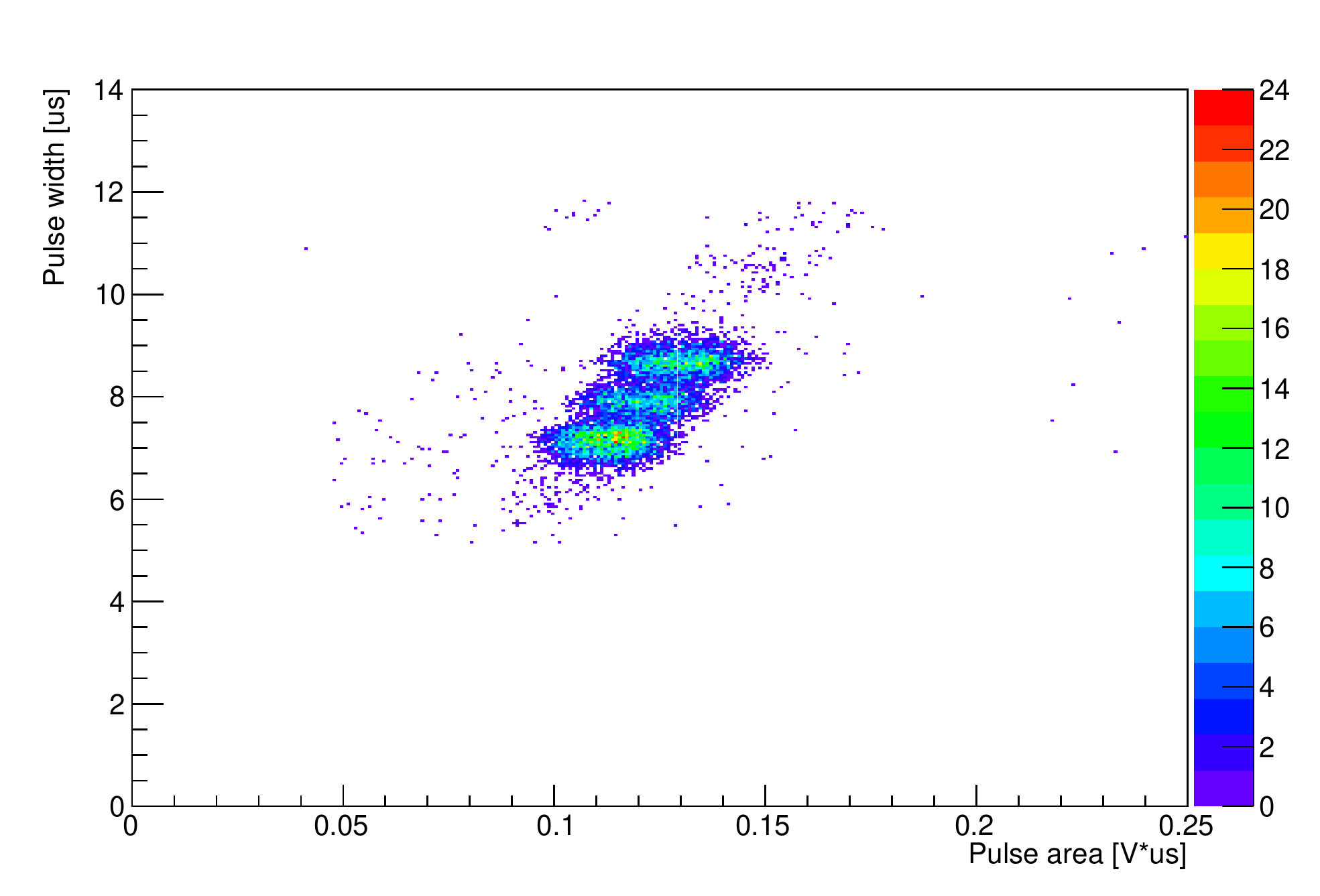}\hspace{2pc}
	\begin{minipage}[b]{14pc}
		\caption{2D plot of pulse width versus pulse area for alpha particles from $^{233}$U (4.8 MeV), $^{239}$Pu (5.2 MeV) and $^{238}$Pu (5.5 MeV) sources, measured in the low-pressure TPC in isobutane at 120 Torr and THGEM gain of 40.}
		\label{2D_PlotTPC}
	\end{minipage}		
\end{figure}

\begin{figure}[h]
	\begin{minipage}{18pc}
		\includegraphics[width=20pc]{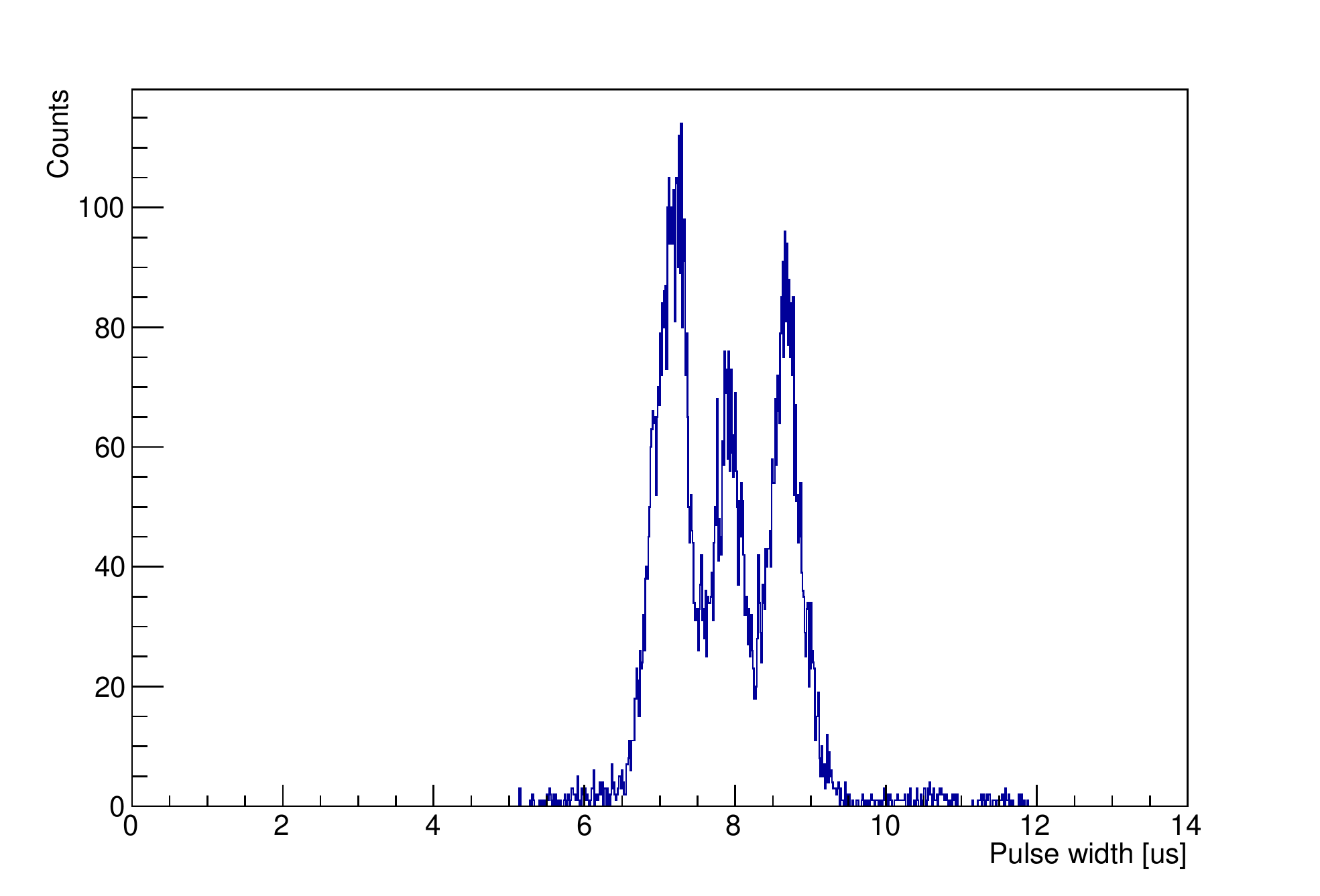}
		\caption{Pulse width spectra of alpha particles from $^{233}$U (4.8 MeV), $^{239}$Pu (5.2 MeV) and $^{238}$Pu (5.5 MeV) sources, measured in the  low-pressure TPC in isobutane at 120 Torr and THGEM gain of 40.}
		\label{TripleSpectraTPC}
	\end{minipage}\hspace{2pc}
	\begin{minipage}{18pc}
		\includegraphics[width=20pc]{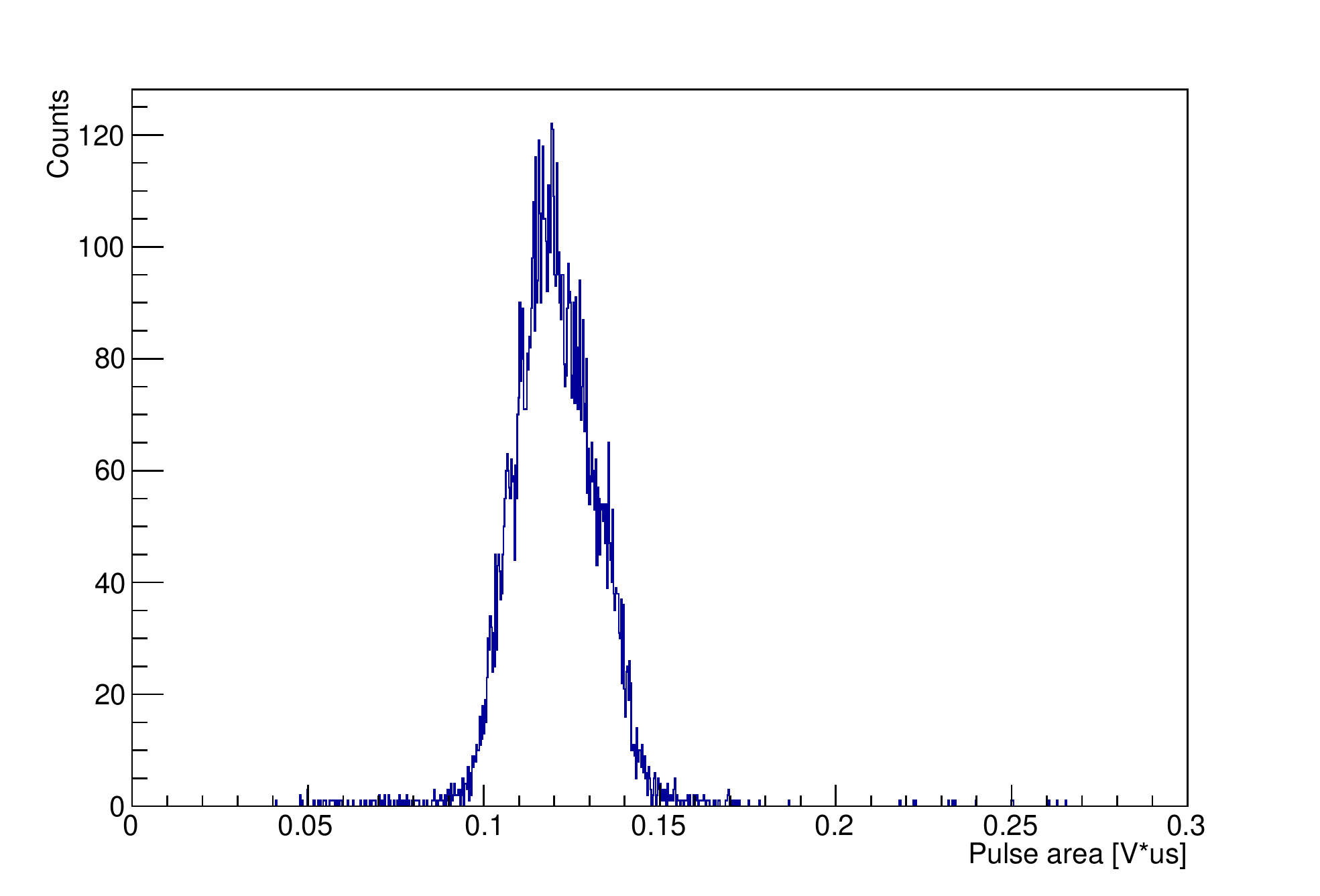}
		\caption{Pulse area spectra of alpha particles from $^{233}$U (4.8 MeV), $^{239}$Pu (5.2 MeV) and $^{238}$Pu (5.5 MeV) sources, measured in the low-pressure TPC in isobutane at 120 Torr and THGEM gain of 40.}	
		\label{TripleSpectraArea}	
	\end{minipage} 
\end{figure}

The performance of the low-pressure TPC with THGEM readout is illustrated in Figs.~\ref{Signal},\ref{2D_PlotTPC},\ref{TripleSpectraArea} and \ref{TripleSpectraTPC}. 
Fig.\ref{Signal} shows the typical pulse shape of the signal from alpha particle. The pulse width is proportional to the alpha particle track range and the pulse area is proportional to the alpha particle energy. 

The measured 2D distribution of the pulse width vs pulse area at a pressure of 120 Torr and THGEM gain of 40 is shown in Fig.~\ref{2D_PlotTPC}. One can distinctly identify three regions corresponding to three alpha particle lines. These regions are well reflected in the projection of the 2D plot on the pulse width axis shown in Fig.~\ref{TripleSpectraTPC}: one can easily distinguish three peaks from alpha-particles of different track ranges. On the other hand, the measurement of the pulse area does not allow to identify these peaks, apparently due to relatively low energy resolution of the THGEM: see Fig.~\ref{TripleSpectraArea}.

High separation capability of this method is illustrated in Table~\ref{tab:res} for two different shaping times of the shaping amplifier.

\begin{table}[h]
	\caption{}
	\label{tab:res}
	\begin{center}
		\begin{tabular}{lllll}
			\br
			Source & Amplifier shaping time & THGEM gain & Pressure & Sigma/Range\\
			\mr
			Triple isotope & 500 ns & 40 & 120 Torr & 3.2\% \\
			Triple isotope & 200 ns & 40 & 120 Torr & 2.2\% \\
			\br
		\end{tabular}
	\end{center}
\end{table}

For Be and B ions we can always reproduce the resolution obtained
for alpha particles, adjusting the track range and E/P values to those
used for alpha particles  by varying the pressure and electric field.

Comparing Figs.~\ref{TripleSourceRange} and ~\ref{TripleSpectraTPC} one can notice the worse experimental resolution compared to the SRIM simulation. This is because the simulation does not take into account the diffusion of the electrons during their drift in the gas volume. The effect of diffusion is clearly seen from the relatively shallow slope of the rear edge of the pulse shape in Fig.~\ref{Signal}. It should be noted that during the joint operation of the chamber with the AMS, an independent start signal (a trigger) will be provided. This will eliminate of the time uncertainly in measuring the rear edge of the pulse, correspondingly improve the time resolution of the low-pressure TPC.  

\section{Conclusions}

A low-pressure TPC with THGEM readout was developed and successfully tested in our laboratory. The track ranges of alpha particles were measured in the TPC with a high accuracy, reaching the 2\,\% precision level. On the basis of these results and the SRIM code simulations, one may conclude that the isobaric boron and beryllium ions (having a range difference of 32\,\%) can be effectively separated coupling the TPC to AMS,  at a level exceeding 10 sigma, by measuring the ion track ranges.

It is expected that this technique will be applied in the AMS facility in Novosibirsk for dating geological objects, in particular for geochronology of Cenozoic Era.

\end{document}